\documentclass[prd,showpacs,twocolumn,preprintnumbers,nofootinbib]{revtex4}

\usepackage[OT2,OT1]{fontenc}
\usepackage{amsmath}
\usepackage{amsfonts}
\usepackage{amssymb}
\usepackage{graphicx}
\usepackage{bm}
\newcommand\be{\begin{equation}}
\newcommand\ee{\end{equation}}
\newcommand\ba{\begin{eqnarray}}
\newcommand\ea{\end{eqnarray}}
\newcommand{\bma}{\left(\begin{array}}
\newcommand{\ema}{\end{array}\right)}

\def\rme{e}
\def\rmd{d}
\def\rmi{i}
\def\p{\partial}

\def\cO{{\cal O}}

\def\cL{{\cal L}}
\def\cM{{\cal M}}
\def\N{\nabla}

\def\a{\alpha}
\def\k{\kappa}
\def\e{\epsilon}
\def\b{\beta}
\def\s{\sigma}

\def\O{\Omega}
\def\bg{\bar{g}}
\def\bR{\bar{R}}
\def\bp{\bar{\partial}}

\def\bN{\bar{\nabla}}
\def\bD{\bar{\Delta}}

\newcommand{\Eq}[1]{(\ref{#1})}

\begin{document}

\title{Detailed balance in Ho\v{r}ava--Lifshitz gravity}
\author{Gianluca Calcagni}
\affiliation{Institute for Gravitation and the Cosmos, Department of Physics,\\ The Pennsylvania State University, 104 Davey Lab, University Park, Pennsylvania 16802, USA}

\date{May 22, 2009}
\begin{abstract}
We study Ho\v{r}ava--Lifshitz gravity in the presence of a scalar field. When the detailed balance condition is implemented, a new term in the gravitational sector is added in order to maintain ultraviolet stability. The four-dimensional theory is of a scalar-tensor type with a positive cosmological constant and gravity is nonminimally coupled with the scalar and its gradient terms. The scalar field has a double-well potential and, if required to play the role of the inflation, can produce a scale-invariant spectrum. The total action is rather complicated and there is no analog of the Einstein frame where Lorentz invariance is recovered in the infrared. For these reasons it may be necessary to abandon detailed balance. We comment on open problems and future directions in anisotropic critical models of gravity.
\end{abstract}

\preprint{arXiv:0905.3740 \hspace{2cm} IGC-09/05-1}
\preprint{PHYSICAL REVIEW D {\bf 81}, 044006 (2010)}\hfill

\maketitle



\section{Introduction}

Following the construction of a Lorentz-violating theory of membranes \cite{Hor1}, Ho\v{r}ava proposed a model of pure gravity which is power-counting renormalizable \cite{Hor2,Hor3}. This has received considerable attention, in particular, regarding its ultraviolet (UV) and infrared (IR) properties \cite{Vis09,Vol09,SVW1,CH,OR,CNPS,SVW2,KLM}, cosmology  \cite{SVW2,TaS,Lif1,KK,Muk09,Bra09,Pia09,Gao09,MNTY,CHZ,Gao2,KeS,KR,CPT,Sar09,Muk2}, spherically symmetric solutions and black holes \cite{KeS,DT,LMP,CCO1,MyK,Myu09,CJ1,Kon09,CJ2,Pan09,BBP,CCO2,Man09}, $pp$-wave and toroidal vacuum solutions \cite{Nas09,Gho09}, vector field configurations \cite{CY}, canonical structure, quantization and unitarity \cite{Nik09,Iza09,Nis09,LiP}, and other issues \cite{ChW}, while inspiring other critical $Dp$-brane or gravitational theories \cite{Klu1,CLS,Klu2}.

In order to reduce the number of operators in the action and simplify some properties of the quantum system, the \emph{detailed balance condition} \cite{Car96} was invoked in \cite{Hor2} (see also \cite{OR,Nis09}). The total ``potential'' term $\cL_V$ in the $D+1$ action descends from a $D$-dimensional Riemannian theory, which is topological massive gravity for $D=3$. This assumption was shown to lead to trouble in the presence of a scalar sector \cite{Lif1}, inasmuch as the signs of the highest-order ($\p^6$) terms in spatial derivatives are opposite. As a consequence, matter is not UV stable. If the scalar field is regarded as the inflaton, its spectrum is not scale invariant. Fixing the sign of the total potential as in the original paper, the tensor spectrum is scale invariant \cite{TaS,Lif1}.\footnote{A negative cosmological constant features in the original proposal but this is not an issue \emph{per se} from the point of view of cosmological observations. Obviously, a massive interacting scalar field can lift the anti-de Sitter vacuum, although later we shall see that this is not the case.}

There are several ways out of the problem. One is to rely on a scale-invariant mechanism different from standard inflation, which exploits the natural freedom from big-bang singularity of the model \cite{Lif1,KK}, and evolve perturbations through a bounce \cite{Bra09}. Another is to notice that, when the parameter $\lambda$ (see below) is different from 1, the pure gravitational theory already contains a dynamical scalar degree of freedom \cite{Hor2,SVW2,KLM,CHZ,CPT}, which has at most 4th-order spatial derivatives if detailed balance holds. Beside the issue of Lorentz invariance for $\lambda\neq 1$, this trace field does not propagate in Minkowski and its physical interpretation has not yet been fully assessed.

A third option, which indeed yields a scale-invariant scalar spectrum \cite{Lif1,Muk09} and was the implicit assumption in many of the above works, is to abandon the detailed balance principle. In either case the role of detailed balance is unclear, especially because only a simplified version of it was implemented in \cite{Lif1}, where the scalar and tensor potentials were decoupled.

It is our purpose to remove this simplification and study the consequences of detailed balance in a model with matter, which is a natural and more realistic extension of the purely gravitational scenario. A scalar theory is the simplest matter field theory one can consider, so that will be our choice. In this sense, this is a most natural follow up of the same study begun in \cite{Lif1}. Moreover, an instability in the inflationary spectrum is an UV instability of the field on a cosmological background, regardless its identification with the inflaton. At least on that particular background, detailed balance may be incompatible with scalar matter and it is desirable to explore the issue of its viability in more detail. For at least all these reasons, it is a legitimate question to ask what are the properties of a scalar in this theory independently from the above-mentioned cosmology-related considerations, which will play little or no role in the present paper.

Although the detailed balance condition is optional, we want to clarify whether, how, and why it constitutes a liability rather than an asset. We shall enforce it to the best of our knowledge and check (i) ultraviolet finiteness; (ii) absence of classical instabilities, e.g., ghosts or scalar potentials unbounded from below; (iii) good infrared limit. To fulfil these conditions in the presence of detailed balance, the total $3+1$ action will need to be defined with a particular choice of sign in front of $\cL_V$. This is the first point of departure with respect to \cite{Hor2}. In doing so, we introduce the operator
\be\label{oper0}
g^{ij}\Delta^{1/2}R_{ij}\,,
\ee
which is expected anyway because pseudodifferential operators already appear in the scalar sector. To show also the effect of a further generalization on dispersion relations, we will actually consider the case
\be\label{operf}
\phi\,g^{ij}\Delta^{1/2}R_{ij}\,.
\ee
Other possibilities shall be taken into account in Sec.~\ref{gene}. The net result is a $3+1$ action with nonminimal couplings, which may realize scale-invariant inflation of trans-Planckian type driven by a scalar field with double-well potential. Interestingly, this is a novel type of scalar-tensor theory which can never be mapped onto an ``Einstein frame'' where matter is decoupled from gravity. Its properties under local conformal transformations of the four-metric are discussed and it is shown that only a particular class of Weyl transformations leads to a minimally coupled IR limit. Outside this class, one would obtain a scenario with varying effective speed of light. In the original Jordan frame, the coupling in the UV limit is minimal only for homogeneous configurations. Notice that nonminimal coupling stems from the prescription for detailed balance, even when the three-dimensional action is minimally coupled [Eq.~\Eq{oper0}].

We find it difficult to obtain a stable Lorentz-invariant limit with a dynamical scalar field. Apart from the amusing possibility to forbid fundamental scalars in the theory, this suggests to abandon detailed balance, in accordance with previous results pointing towards the same conclusion \cite{CNPS,Lif1,LMP,Nas09}. We shall comment on these in the last section.

We wish to stress that the $z=3$ model under consideration is rather general. The nonminimal coupling arises because one is imposing detailed balance together with UV stability. The latter requirement also determines the sign of the total potential. The signs of the coefficients in the three-dimensional action will be arbitrary but this is not the case for the most important coefficients of the $4D$ action, since they are the square of $3D$ couplings (see below). These coefficients appear in a set of constraints which should be satisfied simultaneously. Such rigid conditions are not met generally because of detailed balance. This negative result is insensitive of other details: The instabilities of the theory cannot be adjusted by a different choice of (i) operators, (ii) ordering of derivatives, (iii) the sign of the couplings, or even (iv) the form of the scalar potential. The main problems encountered in this large class of scalar-tensor theories with detailed balance on foliated manifolds are always: (a) absence of a de Sitter vacuum, (b) presence of semiclassical instabilities due to an incompatibility between the sign of the particle (graviton and scalar) kinetic terms and the reality condition on the effective speed of light, and (c) absence of an Einstein frame.

Some properties of pseudodifferential operators are reviewed in Sec.~\ref{pse}. We motivate detailed balance in Sec.~\ref{det}, and construct the gravitational-matter action in Sec.~\ref{action}, where its properties under Weyl transformations are also described. Some generalizations of the simplest model are considered in Sec.~\ref{gene}. Main results, pending problems and future directions are discussed in the last section.


\section{Pseudodifferential operator $\Delta^\a$ and fractional calculus}\label{pse}

Fractional calculus is as old a branch of mathematics as ordinary differential analysis \cite{Kir93,SKM93,Pod99,Mis08,Das08,Jum09}. Its applications range from statistics and long-memory processes such as weather and stochastic financial models \cite{Mis08} to system modeling and control in engineering \cite{Das08}. We shall be interested in fractional powers of the Beltrami--Laplace operator on a manifold $\cM$, which is a particular pseudodifferential elliptic operator \cite{See67,Hor68,Hor71,Sch91,Shu01}.

Let $\zeta$ be the space of entire analytic test functions $\hat\varphi_k$, where $k\in\mathbb{C}$. The space of ultradistributions is the dual of $\zeta$ and its elements are the linear functionals \cite{BBOR}
\be
\hat\psi[\hat\varphi]=\int_\Gamma \rmd k\,\hat\psi_k\hat\varphi_k\,,
\ee
where $\hat\psi_k$ is analytic in $\{k:|{\rm Im}k|>\varrho\}$ and $\hat\psi_k/k^\varrho$ is bounded and continuous on the same domain with $|{\rm Im}k|=\varrho$ included. $\Gamma$ is a contour on the complex plane running clockwise along $|{\rm Im}k|>\varrho$. In coordinate space,
\be
\psi[\varphi]=\int_{-\infty}^{+\infty} \rmd x\, \psi(x)\varphi(x),
\ee
where
\be\label{ft}
\psi(x)=\int_\Gamma \rmd k\,\hat\psi_k\rme^{\rmi kx}\,.
\ee
These equations allow one to define the action of pseudodifferential operators with certain analytic properties, and can be generalized to $D$ spatial dimensions \cite{BBOR} and curved spacetime. For our purposes, it will be sufficient to consider the Fourier transform \Eq{ft} and the pseudodifferential operator
\be
\Delta^\a=[g_{ij}(t,{\bf x})\N^i\N^j]^\a\,,
\ee
where $\a\in \mathbb{R}$ (but in general it can be complex) and $i,j=1,\dots,D$. Here, $D$ is an arbitrary positive integer. The time dependence does not play any role in what follows, as we are interested in single leaves of the spacetime foliation.

The Euclidean and Lorentzian cases were considered in \cite{BBOR,BG}. The Green function of $\Delta^\a$ can be calculated with a number of techniques. For a $D$-dimensional Euclidean space, one obtains
\be
G(|{\bf x}|)\sim |{\bf x}|^{2\a-D}.
\ee
For $\a=1$ this is the usual Newtonian potential: in three dimensions, $G(|{\bf x}|)\sim 1/|{\bf x}|$. When $\a=D/2$, $G(|{\bf x}|)\sim \ln |{\bf x}|$, while for $\a=3$ one has $G(|{\bf x}|)\sim |{\bf x}|^3$. This is the UV correction one expects in Ho\v{r}ava--Lifshitz gravity.

In general, for any $\psi$ 
\be
\Delta^{\a+\b}\psi\neq \Delta^{\a}[\Delta^{\b}\psi]\,,
\ee
unless either $\a$ or $\b$ is natural. Also, given two scalars $A$ and $B$ living in a suitable functional space, the usual integration by parts holds
\ba
\int \rmd^3x\, \sqrt{g}\, A \Delta^\a B&=& \int \rmd^3x\, \sqrt{g}\, (\Delta^\a A)B\nonumber\\
&&+~{\rm boundary~terms}\,,\quad 0\leq\a\leq 1\,,\nonumber\\\label{ab}
\ea
as one can check in momentum space, for both Euclidean space and general Riemannian manifolds \cite{SKM93,Mis08,Shu01,GM,Gua06}.\footnote{The question is whether arbitrary powers of the Beltrami--Laplace operator are self adjoint with respect to the natural $L_2$ scalar product. If the Beltrami--Laplace operator is defined as a self adjoint operator with spectrum in $[0,\infty)$, then one can define its $\a$th fractional power by spectral theory (for $\a>0$), and this is self adjoint.}


In order to compute the total action, one needs the functional variation of Eq.~\Eq{ab} with respect to the metric $g_{ij}$:
\be\label{vd}
A\frac{\delta\Delta^\a}{\delta g_{ij}}B\,.
\ee
When $\a=n$ is a natural number, one obtains a finite sum of contributions
\ba
A(\delta\Delta^n)B&=&\sum_{l=0}^{n-1} A\Delta^l(\delta\Delta)\Delta^{n-1-l} B\nonumber\\
&\to&\sum_{l=0}^{n-1} (\Delta^l A)(\delta\Delta)\Delta^{n-1-l} B\,,\label{nn1}
\ea
where in the last step we have integrated by parts inside the $D$-dimensional integral. For arbitrary $\a$, this expression cannot be readily generalized and the problem of a suitable definition of $\Delta^\a$ arises. To this aim, we can employ the following trick. We first define
\be
\Delta^\a\equiv \rme^{\a\ln\Delta}\,,
\ee
then consider the operator identity \cite{Yan02}
\be
\delta e^{\a X}=\int_0^\a \rmd s\, \rme^{sX}(\delta X)\rme^{(\a-s)X}\,,
\ee
which yields
\be
A(\delta \Delta^\a)B = \int_0^\a \rmd s\, (\Delta^s A)(\delta\ln\Delta)\Delta^{\a-s}B\,,\label{nn2}
\ee
where equality is valid under integration by parts in ${\bf x}$. One can show that Eq.~\Eq{nn2} is equivalent to Eq.~\Eq{nn1} for $\a=n\in\mathbb{N}$.

The logarithm of an operator \cite{Yos73} (well defined as long as the kernel of $\Delta$ is trivial) and its variation can be computed with Borel functional calculus. Other representations of $\delta\Delta^\a$ are possible but, fortunately, in this paper we do not have to enter into detail on the subject. In fact, we will argue that terms of the form \Eq{vd} do not contribute in a way which affects the main UV and IR properties of the total action.

\section{Detailed balance}\label{det}

Let $\cM=\mathbb{R}\times\Sigma$ be a time-space manifold with signature $({-},{+},{+},{+})$ embedding a torsion-free three-dimensional space $\Sigma$ with metric $g_{ij}$, where Latin indices run from 1 to 3. We specialize to three spatial dimensions although most of what will be said can be fairly generalized. On $\Sigma$, we define the space-covariant derivative on a covector $v_i$ as $\N_i v_j \equiv \p_i v_j-\Gamma^l_{ij}v_l$, where $\Gamma^l_{ij}\equiv g^{lm}\left[\p_{(i} g_{j)m}-\tfrac12\p_m g_{ij}\right]$ is the spatial Christoffel symbol. Round brackets denote symmetrized indices, $X_{(ij)}=\left(X_{ij}+X_{ji}\right)/2$. The curvature invariants (under spatial diffeomorphisms) quadratic in spatial derivatives of the metric are the Riemann tensor $R^l_{~imj}\equiv \p_m \Gamma^l_{ij}-\p_j \Gamma^l_{im}+\Gamma^n_{ij}\Gamma^l_{mn}-\Gamma^n_{im}\Gamma^l_{jn}$, the Ricci tensor $R_{ij}\equiv R^l_{~ilj}$, and the Ricci scalar $R\equiv R_{ij}g^{ij}$.

Let $g$ be the determinant of the 3-metric and $\mathbb{G}$ the ``metric of fields'' incorporating both the scalar-field component and the generalized DeWitt metric of metrics \cite{Hor1,Hor2}
\be
{\cal G}_{ijlm}\equiv g_{i(l}g_{m)j}-\frac{\lambda}{3\lambda-1}g_{ij}g_{lm}\,,
\ee
whose inverse is
\be
{\cal G}^{ijlm}\equiv g^{i(l}g^{m)j}-\lambda g^{ij}g^{lm}\,.
\ee
We choose a diagonal metric field
\be\label{met}
\mathbb{G}=\frac12\bma{cc} \k^2{\cal G}^{ijlm} &\,\, 0 \\ 0 &\,\, 1\ema\,,
\ee
which is the usual Wheeler--DeWitt metric in the presence of a scalar field when $\lambda=1$ (e.g., \cite{Wil01}). One also defines the fields
\be
q=\bma{cc} g^{ij}& 0 \\ 0 & \phi\ema\,,\qquad \Pi=\bma{cc} \pi_{ij} & 0 \\ 0 & \pi_\phi\ema\,,
\ee
where the Arnowitt--Deser--Misner momenta are
\ba
\pi_{ij}&=&\frac{\delta S}{\delta \dot g^{ij}}\equiv\frac{2}{\k^2}\sqrt{g}\,{\cal G}_{ijkl}K^{kl}\,,\\
\pi_\phi &=&\frac{\delta S}{\delta \dot \phi}\equiv \sqrt{g}\frac{\dot\Phi}{N}\,,
\ea
and $S$ is the total action. Here, $K_{ij}=K_{ij}(t,{\bf x})$ is the extrinsic curvature
\be\label{K}
K_{ij}=\frac1N \left[\frac12\dot g_{ij}-\N_{(i}N_{j)}\right]\,,
\ee
$N=N(t)$ and $N_i=N_i(t,{\bf x})$ are gauge fields and $\dot{\Phi}\equiv\dot\phi-N^i\p_i\phi$.

We introduce a modification of the Ho\v{r}ava--Lifshitz $3+1$ action with $z=3$ \cite{Hor2} and a matter sector with the following properties: It is (i) invariant under foliated diffeomorphisms, (ii) constructed under the principle of detailed balance, and (iii) nontrivial at the $z=3$ critical point. A fourth property, namely, stability and Lorentz invariance in the infrared, will be checked \emph{a posteriori}.

Consider the total action
\be\label{act}
S=\int_\cM\rmd t\rmd^3x\, \sqrt{g}\,N(\cL_K+\cL_V)\,.
\ee
The kinetic term is
\ba
\cL_K &\equiv& \frac{1}{g}{\rm tr}\left(\Pi\mathbb{G}\Pi\right)\\
&=&\frac{2}{\k^2}\left(K_{ij}K^{ij}-\lambda K^2\right)+\frac{1}{2}\frac{\dot{\Phi}^2}{N^2}\,,\label{kin}
\ea
where tr is the trace and $K\equiv K_i^{\ i}$. $\k^2$ and $\lambda$ are coupling constants with dimension $[\k^2]=z-3$ and $[\lambda]=0$ (hence both dimensionless at the $z=3$ Lifshitz point). The scaling dimensions of the gauge fields are $[N]=0=[g_{ij}]$, $[N_i]=z-1$, while the scalar field has dimension $[\phi]=(3-z)/2$.

The potential $\cL_V$ is determined by detailed balance and follows, in a precise way, from the gradient flow generated by a three-dimensional Euclidean action $W$:
\ba
\cL_V&\equiv& \frac{1}{g}{\rm tr}\left(\frac{\delta W}{\delta q}\mathbb{G}\frac{\delta W}{\delta q}\right)\label{sb0}\\
&=&\frac{\k^2}{8}T_{ij}{\cal G}^{ijlm}T_{lm}+ \frac{1}{2g}\left(\frac{\delta W}{\delta\phi}\right)^2\,,\label{sb1}
\ea
where
\be
T_{ij}\equiv -\frac{2}{\sqrt{g}}\frac{\delta W}{\delta g^{ij}}
\ee
is the stress-energy tensor of the three-dimensional theory. Contrary to \cite{Hor2}, the overall sign in front of the total potential in Eq.~\Eq{act} is positive (this is equivalent to take the $-$ sign and the Euclideanized action $W\to \rmi W$). The reason to do so will soon become apparent. There is also another slight difference with respect to \cite{Hor2}, since $T_{ij}{\cal G}^{ijlm}T_{lm}\neq T^{ij}{\cal G}_{ijlm}T^{lm}$ (unless $\lambda=2/3$).

The detailed balance condition allows for a simple quantization of the system as the Hamiltonian constraint is quadratic and the renormalization group flow can be recast in terms of first-order equations. The total Hamiltonian is
\ba
H&=&\int \rmd^3 x [\dot{q}\Pi-\sqrt{g}N(\cL_K+\cL_V)]\nonumber\\
&=&\int \rmd^3 x (N{\cal H}+N^i{\cal H}_i)\,,
\ea
where the super-Hamiltonian and supermomentum are
\ba
{\cal H} &=& \frac{1}{\sqrt{g}}{\rm tr}\left(\Pi\mathbb{G}\Pi-\frac{\delta W}{\delta q}\mathbb{G}\frac{\delta W}{\delta q}\right)\\
&=&\frac{1}{2\sqrt{g}}\left[\k^2\pi_{ij}{\cal G}^{ijlm}\pi_{lm}
-\frac{\k^2}{4}gT_{ij}{\cal G}^{ijlm}T_{lm}\vphantom{\left(\frac{\delta W}{\delta\phi}\right)^2}\right.\nonumber\\
&&\left.+\pi_\phi^2-\left(\frac{\delta W}{\delta\phi}\right)^2\right]\,,
\ea
\be
{\cal H}_i = \pi_\phi\p_i\phi-2\N^j\pi_{ij}\,.
\ee
By virtue of the detailed balance condition, the Hamilton--Jacobi formalism is naturally implemented and the classical constraints admit a large class of simple solutions. For instance,
\be\label{sim}
\Pi=\frac{\delta W}{\delta q}
\ee
yields solutions of the Hamiltonian constraint which, imposing ${\cal H}_i\approx0$ weakly, respect the scalar equation of motion and conservation of the stress-energy tensor of the three-dimensional theory. In particular, static solutions are obtained when $T_{ij}=0$ \cite{Hor2,Nis09}. These solutions can be found by inverting Eq.~\Eq{sim} with respect to $\dot{g}_{ij}$ and $\dot\phi$, i.e., solving first-order differential equations. In general, there will be also solutions which do not obey Eq.~\Eq{sim}.

To Eq.~\Eq{sim} there corresponds a class of solutions in the quantum theory. The latter inherits the quantum properties of the three-dimensional Riemannian theory. In particular, the detailed balance structure of $S$ is preserved along the renormalization group flow and, if $W$ is renormalizable, then also the $3+1$ theory will be renormalizable \cite{OR}.

A stronger condition on the total potential may be imposed, namely, that the scalar and gravitational sectors factorize (minimal coupling prescription) \cite{Lif1}. Then one can decompose the three-dimensional action $W=W_g+W_\phi$, and determine $W_g$ and $W_\phi$ separately. The gravitational and scalar components of Eq.~\Eq{sim} would split into
\be\label{13}
\pi_{ij}=\frac{\delta W_g}{\delta g^{ij}}\,,\qquad \pi_\phi=\frac{\delta W_\phi}{\delta \phi}\,.
\ee
This leads to a different theory.\footnote{In fact, coupling constants are mutually dependent in the presence of detailed balance. If the coupling constants were all independent, as in ordinary quantum field theories with no detailed balance condition, then Eq.~\Eq{13}
would be a subset of the theory defined by Eq.~\Eq{sb0}. This is not the case with detailed balance because, given the same $W$, in the total action with Eq.~\Eq{sb0} there appear terms [e.g., of the form $\sim(\delta W_\phi/\delta g^{ij})^2$] whose couplings cannot be switched off
without switching off also those of the simplified theory \Eq{13}.} In the remainder of this paper we shall consider the model given by Eq.~\Eq{sb0}.


\section{Action: UV and IR limits}\label{action}

We choose the boundary action $W$ in Eq.~\Eq{sb1} to be
\ba
W&=&\frac{1}{\nu^2}\int \omega_3(\Gamma)+\mu\int\rmd^3 x \sqrt{g}\,[R+\s_0g^{ij}\phi\Delta^{1/2}R_{ij}\nonumber\\
&&\qquad-2L(\phi)]\,.\label{W}
\ea
Here, $\omega_3$ is the Chern--Simons form in terms of the metric-dependent spin connection and
\be\label{L}
L(\phi)\equiv \Lambda_W+\frac14\left(\s_3\phi\Delta^{3/2}\phi+\s_2\phi\Delta\phi-m\phi^2\right)\,.
\ee
The real constants $\nu$, $\mu$, $s_i\equiv\mu\s_i$, $m$, and $\Lambda_W$ have dimension $[\nu]=0$, $[\mu]=1$, $[s_0]=(z-3)/2$, $[s_i]=z-i$ for $i=3,2$, $[m]=z-1$, and $[\Lambda_W]=2$, respectively. The gravitational and matter parts of $W$ were constructed in \cite{Hor2} (without the $\Delta^{1/2}R$ term) and \cite{Lif1}, respectively. The presence of fractional derivatives in the action \Eq{W} leads to a modification of the particle spectrum of the theory (for instance, the solution of a fractional wave equation is no longer a superposition of plane waves; see \cite{BBOR} and references therein) and particle propagation \cite{BG}. It would be interesting to study the classical properties of the model and its solutions. However, not only does this go beyond the scope of the present investigation, but we shall also argue that there are more urgent issues which will eventually characterize the model as physically unviable. Anyway, we will comment on the relevance of pseudodifferential operators in the last two sections.

Using the variations
\be\nonumber
\delta\sqrt{g}=-\frac12 g_{ij}\sqrt{g}\delta g^{ij}\,,
\ee
\be
g^{ij}\delta R_{ij}=[g_{ij}\Delta-\N_{(i}\N_{j)}]\delta g^{ij}\,,\nonumber
\ee
one obtains
\ba
\frac{1}{\sqrt{g}}\frac{\delta W}{\delta g^{ij}}&=&-\frac{2}{\nu^2}C_{ij}+\mu\left[R_{ij}-\frac12\,g_{ij}R+L(\phi)\,g_{ij}\right]\nonumber\\
&&+s_0\left[\phi\left(\Delta^{1/2}R_{ij}-\frac12g_{ij}g^{lm}\Delta^{1/2}R_{lm}\right)\right.\nonumber\\
&&\qquad\left.+g_{ij}\Delta^{3/2}\phi-\N_i\N_j\Delta^{1/2}\phi\right]\nonumber\\
&& +\frac12(s_2\p_i\phi\p_j\phi+s_3\p_i\phi\p_j\Delta^{1/2}\phi)+\dots,\label{del1}
\ea
\ba
\frac{1}{\sqrt{g}}\frac{\delta W}{\delta\phi}&=& s_0g_{ij}\Delta^{1/2}R^{ij}-(s_3\Delta^{3/2}+s_2\Delta-\mu m)\phi\,,\nonumber\\\label{del2}
\ea
where
\be
C_{ij}\equiv \e_i^{\ lm}\N_l\left(R_{mj}-\frac14g_{mj}R\right)
\ee
is the Cotton tensor \cite{Hor2} and $\e^{ilm}$ is the Levi--Civita symbol. The terms not shown in Eq.~\Eq{del1} stem from Eq.~\Eq{nn2} and will not contribute in the following discussion. Later we will see that in the IR there is a frame problem mainly due to the detailed balance prescription and which does not depend on the details of the total action; so the IR limit is unaffected. On the other hand, Eq.~\Eq{nn2} states that Eq.~\Eq{vd} is a superposition of operators of derivative order equal or greater than $\a$. These contribute in the UV only when their order is $3$ and they are contracted [in the sense of Eq.~\Eq{sb0}] with other operators of the same order. Then they would give rise to operators which vanish in the traceless gauge  and/or on homogeneous backgrounds.\footnote{\label{f4} The first part of the statement refers to operators in $\delta W/\delta g^{ij}$ and $\delta W/\delta\phi$ involving (a) the Ricci scalar or (b) the Ricci tensor with all indices contracted, or (c) the Ricci tensor with free indices $i$ and $j$, but contracted with another variation proportional to $g_{ij}$. The second part can be understood, for instance, in a toy model with $z=2$ and no fractional operators ($\a=1$). Consider the operator $\phi\Delta\phi\to g^{ij}\p_i\phi\p_j\phi$ in $W$. Variation with respect to the metric yields $\p_i\phi\p_j\phi$. The only possible term in the total action which might contribute to the UV dispersion relation of the scalar field is given by the contraction of this operator with itself. The result is of the form $(\p\phi\p\phi)^2$ or, after integration by parts, $\phi(\p\phi\p\phi)\p^2\phi$. Perturbing the action and keeping only $O(\delta\phi^2)$ terms, one would get $\delta\phi(\p\phi\p\phi)_*\p^2\delta\phi$, where $(\cdot)_*$ is evaluated on the background. If this is homogeneous, $(\cdot)_*=0$.}

The total action is given by Eqs.~\Eq{act}, \Eq{kin}, \Eq{sb1}, \Eq{del1}, and \Eq{del2}, and is considerably more complicated than one without detailed balance and with the same symmetry requirements ($z=3$ UV fixed point and foliated diffeomorphism invariance).

In the UV limit of the action we keep only 6th-order spatial derivatives and neglect relevant operators. Up to a total derivative (which we discard for simplicity together with any other boundary term) and making use of the twice-contracted Bianchi identity $2\N^iR_{ij}=\N_jR$, the Cotton-Cotton term can be written as
\be
C_{ij}C^{ij}= \frac18R\Delta R -R_{jl} \Delta R^{jl}+R_{jl} \N_i\N^j R^{il}\,.
\ee
Now we can easily check that the dispersion relations of both tensor and scalar sectors are real. To show this, it is sufficient to perturb the action at second order and look at $\p^6$ terms. For simplicity, we choose a flat homogeneous background $g_{ij}^{(0)}(t)$, $\phi^{(0)}(t)$, so that spatial gradient and $R$ terms can be ignored in the background coefficients of the perturbed equations. The perturbation of the 3-metric $g_{ij}= g_{ij}^{(0)}+h_{ij}$ is interpreted as the graviton in transverse-traceless gauge plus eventually the trace scalar mode. Ignoring the latter, the marginal kinetic term is
\be
\frac{\k^2}{2}\left(s_0^2\phi^2-\frac{4}{\nu^4}\right)h_{ij}\Delta^3 h^{ij}\,,
\ee
while for scalar perturbations the 6th-order term is
\be\label{d3}
\left[\frac{s_3^2}{2}-s_0^2\k^2(2\lambda-1)\right]\delta\phi\Delta^{3}\delta\phi\,.
\ee
In order for UV modes to have real frequency, both coefficients in the above equations must be positive, leading to the conditions
\ba
|\phi|&>&\frac{2}{\nu^2|s_0|}\,,\label{constr}\\
s_3^2&>&2s_0^2\k^2(2\lambda-1)\,.\label{constr2}
\ea
If these conditions were violated, UV modes would be unstable. One cannot introduce a UV cutoff to avoid this kind of instability, because the theory is claimed to be UV complete. 
In the case of Eq.~\Eq{oper0}, these constraints reduce to
\be
\nu^2|s_0|>2\,.\label{constr0}
\ee
Since we defined the total potential with an extra $-$ sign with respect to \cite{Hor2}, the scalar UV modes are stable. In \cite{Lif1}, the scalar sector was unstable due to the opposite sign in Eq.~\Eq{d3} with $s_0=0$; here, this problem has been removed. At the same time, the role of the operator \Eq{oper0} or \Eq{operf} is to make the traceless graviton UV modes stable.

The scalar field in the total action acquires an effective potential
\ba
V(\phi)&=&\frac{3(3\lambda-1)\k^2}{2}\frac{\mu^2m^2}{16}\left\{\phi^4\vphantom{\left(\frac{4\Lambda_W}{m}\right)^2}\right.\nonumber\\
&&\left.-8\left[\frac{\Lambda_W}{m}+\frac{2}{3\k^2(3\lambda-1)}\right]\phi^2+\left(\frac{4\Lambda_W}{m}\right)^2\right\}\,.\nonumber\\\label{potot}
\ea
Contrary to \cite{Hor2}, the cosmological constant is \emph{positive}.\footnote{A positive cosmological constant was also obtained in \cite{LMP} by making an analytic continuation of the original, purely gravitational action. However, that formulation is pathological inasmuch as it changes the sign of the $\p^6$ term, thus making the graviton UV unstable.} In \cite{Lif1}, due to the simplified detailed balance condition the scalar potential was $V(\phi)\propto m^2\phi^2$, but scalar perturbations were unstable. It was not possible to add the term Eq.~\Eq{oper0} or \Eq{operf} and change the sign of the total potential to cure the latter problem because the scalar potential would have become unbounded from below, thus leading to another instability.

Provided $\lambda>1/3$, if the effective mass in square brackets is negative $V$ has one global minimum at $\phi=0$, where $V(0)>0$. However, this is in contrast with Eq.~\Eq{constr}, which is a dynamical constraint on the theory. Therefore we are forced to conclude that the effective mass term is positive and it will not be restrictive to choose 
\be
\phi_*^2\equiv\frac{4\Lambda_W}{m}\geq 0\,.
\ee
In this case, $V$ is a double-well potential with minima at
\be\label{min}
\phi_\pm=\pm\sqrt{\frac{8}{3\k^2(3\lambda-1)}+\phi_*^2}\,.
\ee
However, $V(\phi_\pm)<0$, and the anti-de Sitter vacuum is not lifted. Although one can obtain inflation also in anti-de Sitter, this may be a problem of the model. A solution is to flip the sign of the total potential $\cL_V$ and use two different metric of fields: $\mathbb{G}$ in the kinetic term and a metric $\mathbb{G}'$ in the potential term with opposite signature in the scalar-scalar component. Then one can set $s_0=0$ but would break detailed balance, as the class of solutions \Eq{sim} would no longer exist. This option is legitimate, but as the aim of this paper is to enforce detailed balance in all sectors of the theory, we shall not consider it here.

Even when Eq.~\Eq{constr} is satisfied, one cannot yet conclude that the theory is free from classical instabilities. In the IR limit, the Lagrangian density becomes
\ba
\cL_{\rm IR}&\sim&\frac{2}{\k^2}\left[K_{ij}K^{ij}-\lambda K^2+c^2(\phi)R\right]+\frac12\frac{\dot{\Phi}^2}{N^2}-V(\phi)\nonumber\\
&&+s_2\mu\left[\frac{2c^2(\phi)}{3\k^2\mu^2}-m\right]\phi\Delta\phi+\dots\,,\label{actIR}
\ea
where $\cdots$ stand for other relevant operators and
\be\label{c2}
c^2(\phi) \equiv \frac{3(3\lambda-1)\k^4\mu^2m}{16}\left(\phi_*^2-\phi^2\right)\,.
\ee
It is a general result in scalar-tensor and modified gravity models that the presence of (semi)-classical instabilities and superluminal modes depend on the background dynamics \cite{CDD,KM1,Far06,LN,DH,DS}. The coefficients of the $O(\Delta)$ terms (in the action perturbed up to second order in $h_{ij}$ and $\delta\phi$) represent the square of the propagation speed of the physical degrees of freedom. One is Eq.~\Eq{c2} evaluated on the background, the other is
\be
s_2\mu\left[\frac{2c^2(\phi)}{3\k^2\mu^2}-m\right]=s_2\mu m\left[\frac{(3\lambda-1)\k^2}{8}\left(\phi_*^2-\phi^2\right)-1\right].
\ee
They are positive definite if
\be\label{use1}
\phi_*^2-\phi^2>0
\ee
and
\be\label{use2}
\phi_*^2-\phi^2>\frac{8}{(3\lambda-1)\k^2}\,,\qquad {\rm if} \qquad s_2\mu m>0\,.
\ee
These conditions are strong constraints on the dynamics. However,
\be
\phi_+>\phi_*\,,
\ee
and $c^2>0$ when the scalar field is near the local maximum. In the high-mass regime $\phi_*\ll 1$, the effective speed of light is imaginary or, in other words, the field sits at the local maximum. The conclusion is that the theory with a dynamical scalar field does not possess a stable Lorentz-invariant configuration. If $\phi_*\sim\phi_+\gg 1$, the minima are shifted at infinity, and the only stable configuration is a \textit{constant} field. This would naturally lead to a constant effective speed of light, which is presumably a necessary requirement in order to respect observational bounds on Lorentz invariance.

The total action \Eq{act} defines a peculiar scalar-tensor theory where the scalar field is nonminimally coupled with  spatial curvature invariants. Consequently, a conformal transformation from the Jordan frame 
\be
g_{ij} \equiv \O^2(x)\,\bg_{ij}\,,\qquad N_i=\O^2(x)\bar{N}_i\,,\qquad N=\O^z(x)\bar N\,,
\ee
would never lead to a conventional Einstein frame $\bg_{ij}$ where matter is decoupled from gravity at linear level in the Ricci curvature. It may be instructive to make this statement explicit. Let us define
\be
\cO_i\equiv \p_i\ln\O\,,\qquad \cO_{ij}\equiv \bN_i\bN_j\ln\O\,,\qquad \cO\equiv \bD\ln\O\,,
\ee
where $\bD = \O^2\Delta-\cO_i\bp^i$ and indices are raised and lowered with the Einstein metric. The measure scales as $N\sqrt{g}= \O^{3+z}\bar N\sqrt{\bg}$, while the intrinsic curvature invariants are
\ba
\O^{2}R   &=&\bR-2(2\cO+\cO_i \cO^i)\,,\\
\O^{4}R^2 &=& \bR^2-4\bR(2\cO+\cO_i \cO^i)\nonumber\\
              &&+4[4\cO^2+4\cO \cO_i \cO^i+(\cO_i \cO^i)^2]\,,\label{r2}\\
\O^{4}R_{ij}R^{ij} &=& \bR_{ij}\bR^{ij}+2\bR^{ij}(\cO_i \cO_j-\cO_{ij})\nonumber\\
&&-2\bR (\cO+\cO_i \cO^i)+\cO^{ij}\left(\cO_{ij}-2\cO_i \cO_j\right)\nonumber\\
&&+5\cO^2-\frac92\cO\cO_i \cO^i+2(\cO_i \cO^i)^2\,.\label{rr}
\ea
Overall, the action transforms as
\ba
S&=&\int\rmd t\rmd^3x\, \sqrt{\bg}\,\bar N\{\O^{3-z}\bar\cL_K
\nonumber\\
&&+[c^2(\phi)\O^{1+z}+\O^{z-1}F(\cO_i,\cO)]\bR+\dots\}\,,\label{actt}
\ea
where $\dots$ are all the other operators and $F(\cO_i,\cO)$ can be found from the above transformation of the higher curvature invariants $R^2$ and $R_{ij}R^{ij}$:
\be\label{ef}
F(\cO_i,\cO)=-2\a_2(\cO+\cO_i\cO^i)-4\a_1(2\cO+\cO_i\cO^i)\,,
\ee
where $\a_1$ and $\a_2$ are the (constant) coefficients of the $R^2$ and $R_{ij}R^{ij}$ terms, respectively. In this sense, the conformal transformation does not ``preserve'' the renormalization group flow, as the renormalization group properties of the operators change from one frame to another. 
At the IR point one may define the conformal transformation such that
\be\label{sp}
\O^2[F(\cO_i,\cO)+c^2(\phi)\O^2]={\rm const}\,.
\ee
Then, the linear Ricci scalar part in the IR Lagrangian density is minimally coupled. However, this is true only on inhomogeneous backgrounds: Eq.~\Eq{sp} states that $\O=\O(\phi,\p_i\phi)$, while Eq.~\Eq{ef} is nonzero only if the field $\phi$ is not homogeneous, $\p_i\phi\neq 0$.


\section{Some generalizations}\label{gene}

So far we have assumed a modification of the three-dimensional action $W$ of \cite{Hor2} (topologically massive gravity) and \cite{Lif1} of the form Eq.~\Eq{oper0} or \Eq{operf}. In this section we comment on three possible generalizations:
\begin{itemize}
\item Operators which are nonlinear in the scalar field $\cO_f\sim g_{ij}f(\phi)\Delta^{1/2}R_{ij}$.
\item General (self-interacting) scalar potentials $U(\phi)$.
\item Fractional derivatives in the gravity sector.
\end{itemize}
Besides the fact that an operator of the form 
\be\label{frg}
\cO_f\sim g_{ij}f(\phi)\Delta^{1/2}R_{ij}
\ee
would further complicate an already intolerably cumbersome model, it would not improve or change the above stability analysis in the UV (in the IR, $\cO_f$ is irrelevant and this conclusion is trivial). Actually, the only change would be in the conditions \Eq{constr} and \Eq{constr2}, which would become
\ba
|f(\phi)|&>&\frac{2}{\nu^2|s_0|}\,,\label{constrf}\\
s_3^2&>&2s_0^2\k^2(2\lambda-1)[f'(\phi)]^2\,.\label{constrf2}
\ea
These equations constrain the allowed values of $\phi$ according to the form of the function $f(\phi)$. However, we have seen that the main problems arise in the infrared, so any specific choice for $f$ would be of interest only after addressing the latter. Therefore Eq.~\Eq{oper0} does not lead to any loss of generality. 

A choice one could modify in the infrared is the one for the scalar potential in $W$. In Eq.~\Eq{L}, we picked a quadratic potential because, as we argued in \cite{Lif1}, a more general form would neither modify the physics nor, as we can show with an example, relax the infrared problem. Replace $m\phi^2/4\to U(\phi)$ in Eq.~\Eq{L}. The four-dimensional scalar potential reads as
\be\label{pototu}
V(\phi)\propto U^2-2\Lambda_WU-\frac{4}{3\k^2(3\lambda-1)}{U'}^2+\Lambda_W^2\,,
\ee
where the overall normalization constant is positive if $\lambda>1/3$. Taking $U=a\phi^3$, where $a$ is a constant, the potential is
\be\label{phi3}
V(\phi)\propto \phi^6-\frac{12}{\k^2(3\lambda-1)}\phi^4-\frac{2\Lambda_W}{a}\phi^3+\frac{\Lambda_W^2}{a^2}\,.
\ee
Without any loss of generality, let $\Lambda_W/a>0$. $V$ is bounded from below with a saddle point at $\phi=0$ and a global minimum at some $\phi_+>0$ (if $\Lambda_W/a<0$, the minimum is at $-\phi_+$). If $f\propto \phi^n$, Eqs.~\Eq{constrf} and \Eq{constrf2} place a parameters-dependent constraint $\phi_1(\nu^2,s_0,s_3,n,\dots)<|\phi|<\phi_2(\nu^2,s_0,s_3,n,\dots)$ which can be tuned to include $\phi_+$. For instance, when $f(\phi)=\phi$ we can always fix the coefficients in Eq.~\Eq{phi3} such that $\phi_+>2/(\nu^2|s_0|)$. Nonetheless, the reader can check that the anti-de Sitter problem still holds, as $V(\phi_+)<0$.

In the infrared, the constraint \Eq{use1} on the graviton kinetic term (reality of the effective speed of light $c$) becomes
\be\label{useu1}
U(\phi)<\Lambda_W\,,
\ee
which is never satisfied near the minimum $\phi_+$. When $U=a\phi^3$, Eq.~\Eq{useu1} reads $\phi<\phi_*\equiv(\Lambda_W/a)^{1/3}$, but one can show that $\phi_+>\phi_*$. Hence, $c(\phi)\in\mathbb{R}$ only away from the minimum.

Finally, the gravitational sector could be decorated with other fractional operators apart from Eq.~\Eq{oper0}. The number of these operators is infinite, so it is not possible to write the most general action in a tractable form. Fortunately, we can still say something about most of the operators we have ignored so far. We consider the UV and IR limits separately and in this order.

Near the UV fixed point we are interested only in operators in $W$ of derivative order 3. Operators made of $n$ Ricci tensors or scalars and the $\a$-th power of the Laplacian are constrained to have $\a=(3/2)-n$. Operators with $n>1$ would contribute terms which vanish on flat homogeneous backgrounds, so they would not affect the above stability analysis (of course a choice of other backgrounds is possible but we will not consider it here). On the other hand, operators with $n<1$ would generate terms in the total action which diverge on the same backgrounds; we prefer to avoid this situation, which resembles modified gravity models of type $1/R$. Setting $n=1$, only three possibilities remain: Eq.~\Eq{frg} and
\be
R^{ij}f(\phi)\Delta^{1/2}g_{ij}\,,\qquad f(\phi)\Delta^{1/2}R\,.\nonumber
\ee
Marginal terms yielded by a functional variation of these objects would either vanish in the traceless gauge (compare footnote \ref{f4}) or be equivalent to those given by Eq.~\Eq{frg}. 

The other crucial scale at which new operators might affect the conclusions of the previous sections is near the IR fixed point. Pseudodifferential operators change, sometimes dramatically, the particle spectrum of a theory whenever they make their appearance. Therefore any fractional operator dominating in the infrared is very likely to spoil both general relativity and quantum field theory on sufficiently large scales. Therefore operators of total derivative order smaller than 2 must be excluded not only in the $4D$ action, but also in $W$,
because there is always a constant term in the $3D$ action which would be multiplied by variations of such operators. This category includes matter operators of the form $\phi\Delta^\a\phi$, $\a<1$, which we had already omitted in Eq.~\Eq{L}, plus many others like
\be\nonumber
f(\phi) \Delta^\a R\,,\qquad f(\phi) g^{ij}\Delta^\a R_{ij}\,,\qquad \a<0\,.
\ee
Notice that exclusion of these operators on phenomenological ground is legal as long as one is concerned only with the definition of the total action without worrying about its self consistency along the renormalization group flow. In other words, if a full renormalization analysis showed the necessity of any of the above terms, then they could no longer be excluded. However, they can be generated only if other fractional operators are already present, which is the case only with detailed balance. This would lead to a modification of the IR limit. If this is phenomenologically harmless, then these operators do not affect the detailed balance issue. If, on the other hand, the new IR limit is problematic, then there must be no fractional operators at all and, hence, no detailed balance. Therefore these lowest-order operators cannot possibly improve the predicament of detailed balance. If there was need to further convince the reader, we would also note that the IR frame problem is independent of all these considerations, so it is robust regardless of changes in the action at this or any other scale.

The absence of an Einstein frame in the infrared is not cured by any of the above generalizations. To summarize, the problems stemming from the detailed balance condition seem to be rather model independent and hard to eradicate via reasonable modifications of the three-dimensional action.


\section{Discussion}\label{conc}

The detailed balance condition is not physically necessary but it leads to a class of simple classical and quantum solutions. We asked whether there exists a viable semiclassical model of Ho\v{r}ava--Lifshitz gravity with scalar matter which obeys it. All the above arguments rely on particular backgrounds and ignore gauge issues but the answer appears to be negative and, to some extent, independent of the details of the model.

There is evidence that it is difficult to achieve a good infrared limit in near-homogeneous backgrounds. This may favour simpler theories without detailed balance (e.g., \cite{SVW1,SVW2,Nas09}), at least as far as their classical dynamics is concerned. This is not in contrast with the findings of \cite{CNPS}, where it was argued that detailed balance leads to strong coupling on all scales, thus implying that such a theory (without matter) does not have a perturbative IR limit (however, see \cite{Muk2}); a similar conclusion about an anomalous IR limit in the presence of detailed balance was reached in \cite{LMP}. 

Perhaps one cannot yet draw a positive conclusion regarding the viability of detailed balance, since the definition of the metric $\mathbb{G}$ has a certain degree of arbitrariness and might be extended to nondiagonal \emph{Ans\"atze} which simplify the flow equations.\footnote{This can modify the IR limit but near the UV fixed point the theory is unlikely to change much. Nondiagonal terms are proportional to the product of Eq.~\Eq{del2} times the trace of Eq.~\Eq{del1}. A direct inspection shows that the UV dispersion relation of the graviton is left untouched, while Eq.~\Eq{d3} acquires a contribution proportional to $s_0s_3$, which would just place a constraint similar to Eq.~\Eq{constr2} on the relative magnitude of the couplings $s_0$ and $s_3$.} However, although neither Ho\v{r}ava--Lifshitz gravity with matter nor its renormalization group flow have been studied thoroughly, the above instabilities are intrinsic to the theory. In fact, the existence of a bad IR limit with not even approximate Lorentz invariance (varying speed of light, no Einstein frame, and so on) is a consequence of how the model is structured, i.e., it is due to the detailed balance condition. This condition propagates into the quantum theory \cite{OR}, so the tree-level results are enough to illustrate the possibility of severe fine-tunings in the model.

Whether or not detailed balance is assumed, Ho\v{r}ava--Lifshitz gravity without matter possibly suffers from problems which the presence of matter not only might not cure, but could also aggravate. The respect of observational constraints on Lorentz invariance is not yet guaranteed. So far, the problem of Lorentz invariance has been considered only at tree level in the literature; the dispersion relation for $z=3$ Lifshitz fields is such that the tree-level theory is safely within experimental bounds. However, it was argued in \cite{CPSUV,CPS} that loop corrections to the propagator of fields in a Lorentz-violating theory of quantum gravity generally lead to violations several orders of magnitude larger than the tree-level estimate, unless the bare parameters of the model are fine-tuned.\footnote{Specifically for Ho\v{r}ava--Lifshitz gravity, a possible fine tuning problem was also pointed out in \cite{Vol09}.} This and the above issues will deserve further consideration.

On a positive note, the model presented here is characterized by stimulating technical issues and phenomenology. In particular, fractional operators can lead to particle spectra considerably different from the usual ones. Also, theories with anisotropic scaling are natural settings wherein to implement the popular notion that ``spacetime is fractal'' at high energies and microscopic scales. The critical exponent $z$ determines the spectral dimension of spacetime, which flows from $1+D$ in the infrared to $1+D/z$ in the ultraviolet \cite{Hor3}. In particular, at small scales ($D=z$) these models are two-dimensional, inasmuch as physical degrees of freedom propagate on an effective (possibly fractal) two-dimensional geometry. As we have seen above, the Newton potential does change according to the scale, so that the large-scale behaviour $G(|{\bf x}|)\sim |{\bf x}|^{-1}$ is replaced by $G(|{\bf x}|)\sim |{\bf x}|^{2z-D}$. Since integrals on net fractals (e.g., self-similar or cookie-cutter sets) can be approximated by fractional integrals \cite{RLWQ}, it is natural to consider fractional integrals over a space with fractional dimension; on the associated phase space, one can construct fractional Hamiltonian systems \cite{Tar1,Tar2,Tar3,Tar4,Tar5,Tar6,Tar7}.\footnote{Lagrangian and Hamiltonian mechanics with fractional derivatives \cite{Rie96,Rie97} describe classical systems with nonconservative forces such as friction.} Fractional integral actions (Stieltjes actions) \cite{UO} have applications, for instance, in economics, and
admit a neat geometrical and physical interpretation \cite{Bul88,Pod02}. It would be interesting to develop these ideas in more detail.


\begin{acknowledgments}
The author is supported by NSF Grant No. PHY0854743, the George A. and Margaret M. Downsbrough Endowment and the Eberly research funds of Penn State. He thanks G. Grubb for helpful email correspondence and S. Mukohyama for his initial involvement in this project.
\end{acknowledgments}
 

\end{document}